\documentstyle[multicol,aps,psfig]{revtex}
\renewcommand{\narrowtext}{\begin{multicols}{2}
\global\columnwidth20.5pc\noindent}
\renewcommand{\widetext}{\end{multicols}
\global\columnwidth42.5pc}
\begin{document}
\draft
\preprint{7 October 1999}
\title{Competing Ground States of the New Class of Halogen-Bridged
       Metal Complexes}
\author{Shoji Yamamoto}
\address
{Department of Physics, Okayama University,
 Tsushima, Okayama 700-8530, Japan}
\date{Received 7 October 1999}
\maketitle
\begin{abstract}
Based on a symmetry argument, we study the ground-state properties of
halogen-bridged binuclear metal chain complexes.
We systematically derive commensurate density-wave solutions from a
relevant two-band Peierls-Hubbard model and numerically draw the
the ground-state phase diagram as a function of electron-electron
correlations, electron-phonon interactions, and doping concentration
within the Hartree-Fock approximation.
The competition between two types of charge-density-wave states,
which has recently been reported experimentally, is indeed
demonstrated.
\end{abstract}
\narrowtext

   Halogen-bridged transition-metal linear-chain complexes
(MX chains; ${\rm M}={\rm metal}$, ${\rm X}={\rm halogen}$) have been
a central subject in material science due to their low dimensionality
and competing electron-electron (el-el) and electron-phonon (el-ph)
interactions, which lead to the variety of electronic states.
PtX compounds \cite{Clar95} provided various topics of both chemical
and physical interest, such as intense dichroism and unusual Raman
spectra which originate in their mixed-valence ground states.
NiX compounds \cite{Toft61,Tori41,Okam81} with mono-valence magnetic
ground states visualized the competition between the Peierls and Mott
insulators.
Nasu \cite{Nasu65} gave a theoretical description of the ground-state
competition.
The Los Alamos group \cite{Gamm08} extensively investigated potential
ground states, including incommensurate density waves \cite{Bati28}
and spin-frustrated novel antiferromagnetism \cite{Rode98}.

   In recent years, the new class of these family compounds (MMX
chains), which consists of binuclear metal units bridged by halogen
ions, has been stimulating further interest in this system.
The presence of the metal-metal bonds allows us to expect a wider
variety of electronic structures \cite{Conr23,Kimu40,Kita}.
K$_4$[Pt$_2$(pop)$_4$X]$\cdot$$n$H$_2$O
($\mbox{X}=\mbox{Cl},\mbox{Br},\mbox{I}$;
 $\mbox{pop}=\mbox{P}_2\mbox{O}_5\mbox{H}_2$)
\cite{Che04,Kurm20} and
M$_2$(dta)$_4$I
($\mbox{M}=\mbox{Pt},\mbox{Ni}$; $\mbox{dta}=\mbox{CH}_3\mbox{CS}_2$)
\cite{Bell44}
are both typical of the MMX compounds but their electronic properties
look rather different.
Pt$_2$(dta)$_4$I shows metallic conduction above room temperature
\cite{Kita}, whereas no metallic behavior has ever been found in the
pop series.
Their ground states are both charge density waves
\cite{Conr23,Kimu40,Kita} but show different valence structures.
The variety of ground states must be one of the most interesting
consequences of intrinsic multi-band effects and competing el-el and
el-ph interactions.
Here, based on a symmetry argument \cite{Ozak55,Yama29,Yama22}, we
reveal rich phase diagrams of the one-dimensional (1D) two-band
three-orbital Peierls-Hubbard model.
{\it The competition between the two distinct charge-density-wave
states is indeed demonstrated.
Antiferromagnetic ground states for the Ni complexes and the onset of
further novel phases under hole doping are also predicted.}

   We introduce the 1D model Hamiltonian as
\widetext
\begin{eqnarray}
   {\cal H}
   &=&\sum_{n,s}
      \big[
       (\varepsilon_{\rm M}-\beta\delta_{1:n})n_{1:n,s}
      +(\varepsilon_{\rm M}+\beta\delta_{2:n})n_{2:n,s}
      + \varepsilon_{\rm X}n_{3:n,s}
      \big]
    + \frac{K}{2}\sum_{n}
      \big(
       \delta_{1:n}^2+\delta_{2:n}^2
      \big)
      \nonumber \\
   &-&\sum_{n,s}
      \big[
       (t_{\rm MX}-\alpha\delta_{1:n})a_{1:n,s}^\dagger a_{3:n,s}
      +(t_{\rm MX}+\alpha\delta_{2:n})a_{2:n,s}^\dagger a_{3:n,s}
      + t_{\rm MM}\,a_{1:n,s}^\dagger a_{2:n-1,s}
      + {\rm H.c.}
      \big]
      \nonumber \\
   &+&\sum_{n}
      \big(
       U_{\rm M}\,n_{1:n,+}n_{1:n,-}
      +U_{\rm M}\,n_{2:n,+}n_{2:n,-}
      +U_{\rm X}\,n_{3:n,+}n_{3:n,-}
      \big)
      \nonumber \\
   &+&\sum_{n,s,s'}
      \big(
       V_{\rm MX}\,n_{1:n,s}n_{3:n  ,s'}
      +V_{\rm MX}\,n_{2:n,s}n_{3:n  ,s'}
      +V_{\rm MM}\,n_{1:n,s}n_{2:n-1,s'}
      \big) \,,
   \label{E:H}
\end{eqnarray}
\narrowtext
where $n_{i:n,s}=a_{i:n,s}^\dagger a_{i:n,s}$ with
$a_{i:n,s}^\dagger$ being the creation operator of an electron with
spin $s=\pm$ (up and down) for the M$d_{z^2}$ ($i=1,2$) or X$p_z$
($i=3$) orbital in the $n$th MXM unit, respectively, and
$\delta_{i:n}=u_{3:n}-u_{i:n}$ with $u_{i:n}$ being the
chain-direction displacement of the metal ($i=1,2$) or halogen
($i=3$) ion in the $n$th MXM unit.
$\alpha$ and $\beta$ are, respectively, the intersite and intrasite
el-ph coupling constants, while $K$ is the metal-halogen spring
constant.
We assume, based on experimental observations, that the MM units are
not deformed ($u_{1:n}=u_{3:n-1}$).
$\varepsilon_{\rm M}$ and $\varepsilon_{\rm X}$ are the on-site
energies (electron affinities) of isolated metal and halogen atoms,
respectively.
The electron hoppings between these levels are modeled by
$t_{\rm MM}$ and $t_{\rm MX}$, whereas the el-el Coulomb interactions
by $U_{\rm M}$, $U_{\rm X}$, $V_{\rm MM}$, and $V_{\rm MX}$.
The symmetry group of the system is given by
${\bf G}={\bf P}\times{\bf S}\times{\bf T}$\,,
where ${\bf P}={\bf L}\land{\bf C}_2$ is the space group of a linear
chain with ${\bf L}$ being the 1D translation group whose basis
vector is the unit-cell translation, ${\bf S}$ the group of
spin-rotation, and ${\bf T}$ the group of time reversal.
Let $\check{G}$ denote the irreducible real representations of
{\bf G},
where their representation space is
spanned by the Hermitian operators
\{$a_{i:k,s}^\dagger a_{j:k',s'}$\}.
There is a one-to-one correspondence between $\check{G}$ and
broken-symmetry phases of density-wave type.
Any representation $\check{G}$ is obtained as a Kronecker product of
the irreducible real representations of ${\bf P}$, ${\bf S}$, and
${\bf T}$;
$\check{G}=\check{P}\otimes\check{S}\otimes\check{T}$\,.
$\check{P}$ is characterized by an ordering vector $q$ in the
Brillouin zone and an irreducible representation of its little group
${\bf P}(q)$, and is therefore labeled $q\check{P}(q)$.
We denote the relevant representations of ${\bf S}$ by $\check{S}^0$
(singlet) and $\check{S}^1$ (triplet), while those of ${\bf T}$ by
$\check{T}^0$ (symmetric) and $\check{T}^1$ (antisymmetric).
$\check{P}\otimes\check{S}^0\otimes\check{T}^0$,
$\check{P}\otimes\check{S}^1\otimes\check{T}^1$,
$\check{P}\otimes\check{S}^0\otimes\check{T}^1$, and 
$\check{P}\otimes\check{S}^1\otimes\check{T}^0$,
respectively, describe charge-density-wave (CDW), spin-density-wave
(SDW), charge-current-wave (CCW), and spin-current-wave (SCW) states.
We leave out all the current-wave states, which either result in the
one-way uniform flow or break the charge-conservation law in the
present system.
We consider density waves of two types, $q=0$ and $q=\pi$, which are
labeled $\mit\Gamma$ and $X$, respectively.
Thus the representations labeled
$K\check{P}(K)
 \otimes\check{S}^i\otimes\check{T}^i$
($K={\mit\Gamma},X$; $i=0,1$) are of our interest.
Since ${\bf P}({\mit\Gamma})={\bf P}(X)={\bf C}_2$,
$\check{P}({\mit\Gamma})$ and $\check{P}(X)$ are either
$A$ ($C_2$-symmetric) or $B$ ($C_2$-antisymmetric) representation.

Then the Hartree-Fock (HF) approximation of the Hamiltonian
(\ref{E:H}) in the momentum space is given by
\begin{equation}
   {\cal H}_{\rm HF}
   =\sum_{K={\mit\Gamma},X}
    \sum_{\lambda=0,z}
    \sum_{i,j}\sum_{k,s,s'}
    x_{ij}^{\lambda}(K;k)\,
    a_{i:k+q,s}^\dagger a_{j:k,s'}
    \sigma_{ss'}^\lambda\,.
   \label{E:HHF}
\end{equation}
$x_{ij}^\lambda(K;k)$ is self-consistently given in terms of the
density matrices
$\rho_{ij}^\lambda(K;k)
 =\frac{1}{2}\sum_{s,s'}
  \langle a_{j:k+q,s}^\dagger a_{i:k,s'}\rangle_{\rm HF}\,
  \sigma_{ss'}^\lambda$,
where the $2\times 2$ matrices $\sigma^\lambda$ are the unit
($\lambda=0$) and Pauli ($\lambda=x,y,z$) matrices, and
$\langle\cdots\rangle_{\rm HF}$ means the quantum average in a HF
eigenstate.
We note that no helical-spin ($\lambda=x,y$) solution is obtained
from the present model.
When ${\cal H}_{\rm HF}$ is decomposed into spatial-symmetry-definite
components as
\vskip -1mm
\begin{figure}
\begin{flushleft}
\qquad\quad\ \mbox{\psfig{figure=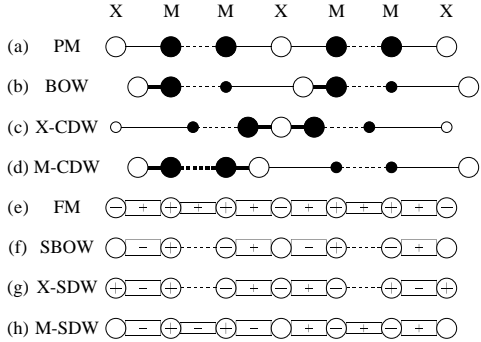,width=70mm,angle=0}}
\end{flushleft}
\vskip 1mm
\caption{Schematic representation of possible density-wave states,
         where the variety of circles and segments qualitatively
         represents the variation of local charge densities and
         bond orders, respectively, whereas the signs $\pm$ in
         circles and strips describe the alternation of local spin
         densities and spin bond orders, respectively.
         Circles shifted from the regular position qualitatively
         represent lattice distortion, which is peculiar to
         nonmagnetic phases.}
\label{F:DW}
\end{figure}
\begin{equation}
  {\cal H}_{\rm HF}
   =\sum_{D=A,B}\sum_{K={\mit\Gamma},X}\sum_{\lambda=0,z}
    h^\lambda(K;D)\,,
\end{equation}
the broken-symmetry Hamiltonian for the representation
$KD\otimes\check{S}^i\otimes\check{T}^i$ is given by
$h^0({\mit\Gamma};A)+h^\lambda(K;D)$, where $\lambda=0$ for $i=0$ and
$\lambda=z$ for $i=1$.
The charge ($\lambda=0$) and spin ($\lambda=z$) densities on site
$i$ in the $n$th MXM unit are expressed as
$
   d_{i:n}^\lambda
     =\frac{1}{2}\sum_{s,s'}
     \langle a_{i:n,s}^\dagger a_{i:n,s'}\rangle_{\rm HF}
     \sigma_{ss'}^\lambda
$,
while the bond ($\lambda=0$) and spin bond ($\lambda=z$) orders
between site $i$ in the $n$th MXM unit and site $j$ in the $m$th MXM
unit are defined as
$
   p_{i:n;j:m}^\lambda
     =\frac{1}{2}\sum_{s,s'}
     \langle a_{i:n,s}^\dagger a_{j:n,s'}\rangle_{\rm HF}
     \sigma_{ss'}^\lambda
$.
The lattice displacements $u_{i:n}$ are self-consistently determined
so as to minimize the HF energy
$E_{\rm HF}\equiv\langle{\cal H}\rangle_{\rm HF}$.
Further group-theoretical procedure will be detailed elsewhere.
Here we briefly characterize all the phases and schematically show
them in Fig. \ref{F:DW}.
\begin{enumerate}
\renewcommand{\labelenumi}{(\alph{enumi})}
\item ${\mit\Gamma}A\otimes\check{S}^0\otimes\check{T}^0$:
    The paramagnetic state with the full symmetry {\bf G},
    abbreviated as PM.

\item ${\mit\Gamma}B\otimes\check{S}^0\otimes\check{T}^0$:
    Bond order wave with polarized charge densities in the M
    sublattice, abbreviated as BOW.

\item $XA           \otimes\check{S}^0\otimes\check{T}^0$:
    Charge density wave on the X sublattice with alternating
    polarized charge densities in the M sublattice, abbreviated as
    X-CDW.

\item $XB           \otimes\check{S}^0\otimes\check{T}^0$:
    Charge density wave on the M sublattice, abbreviated as M-CDW.

\item ${\mit\Gamma}A\otimes\check{S}^1\otimes\check{T}^1$:
    Ferromagnetism with uniform spin bond orders, abbreviated as FM.

\item ${\mit\Gamma}B\otimes\check{S}^1\otimes\check{T}^1$:
    Spin bond order wave with polarized spin densities in the M
    sublattice, abbreviated as SBOW.

\item $XA           \otimes\check{S}^1\otimes\check{T}^1$:
    Spin density wave on the X sublattice with alternating polarized
    spin densities in the M sublattice, abbreviated as X-SDW.

\item $XB           \otimes\check{S}^1\otimes\check{T}^1$:
    Spin density wave on the M sublattice, abbreviated as M-SDW.
\end{enumerate}
The nonmagnetic phases (a)-(d), respectively, correspond to
the averaged-valence, charge-polarization, alternating
charge-polarization, and CDW states in the experimental convention
\cite{Kita}, the latter three of which exhibit el-ph coupling.
X-CDW is a novel density-wave state stabilized by M-sublattice
distortion.
It is the case for the MX-type complexes that the metal ions are
locked in the surrounding ligands.
Although the binuclear metal units are still heavily surrounded by
ligands, $^{129}$I M\"ossbauer spectroscopy measurements \cite{Kita}
for Pt$_2$(dta)$_4$I imply the ground state of the X-CDW type.
M-sublattice distortion has not yet been reported for the pop series,
where the ground state of the M-CDW type appears \cite{Kimu40}.
This competition is a central issue in our argument.
On the other hand, the CDW-SDW competition is another interesting
issue.
Magnetic instabilities are generally not coupled with phonons.
\widetext
\vskip 0mm
\begin{figure}
\begin{flushleft}
\qquad\qquad\quad\ 
\mbox{\psfig{figure=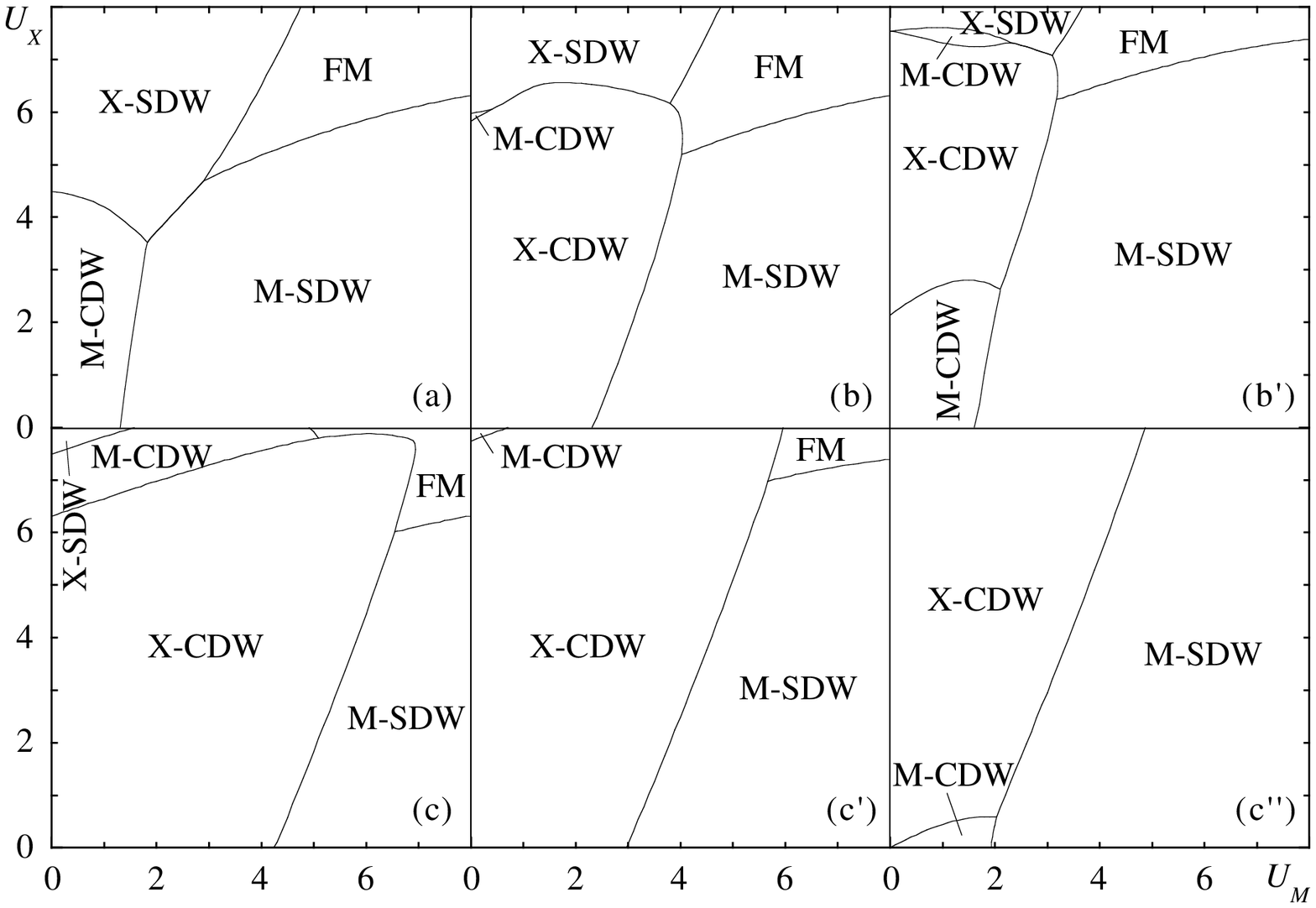,width=160mm,angle=0}}
\end{flushleft}
\vskip 0mm
\caption{Typical ground-state phase diagrams at $\frac{5}{6}$ band
         filling, where the parametrization,
         $V_{\rm MM}=V_{\rm MX}=1.0$, is common to all.
         (a) $\alpha=0.9$, $\beta=0.6$, 
             $\varepsilon_{\rm M}-\varepsilon_{\rm X}=1.0$.
         (b) $\alpha=1.2$, $\beta=0.6$, 
             $\varepsilon_{\rm M}-\varepsilon_{\rm X}=1.0$;
         (b$'$) $\alpha=1.2$, $\beta=0.6$, 
             $\varepsilon_{\rm M}-\varepsilon_{\rm X}=2.0$.
         (c) $\alpha=1.5$, $\beta=0.6$, 
             $\varepsilon_{\rm M}-\varepsilon_{\rm X}=1.0$;
         (c$'$) $\alpha=1.5$, $\beta=0.6$, 
             $\varepsilon_{\rm M}-\varepsilon_{\rm X}=2.0$;
         (c$''$) $\alpha=1.5$, $\beta=0.6$, 
             $\varepsilon_{\rm M}-\varepsilon_{\rm X}=3.0$.}
\label{F:PhD5}
\end{figure}
\narrowtext
\vspace*{1mm}
\begin{figure}
\begin{flushleft}
\qquad\quad\ \mbox{\psfig{figure=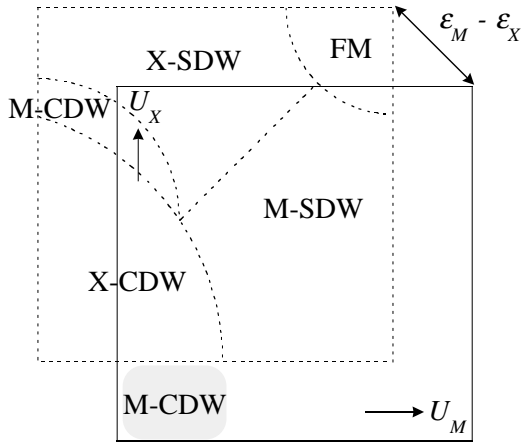,width=70mm,angle=0}}
\end{flushleft}
\vskip 2mm
\caption{Qualitative understanding of the phase diagrams at
         $\frac{5}{6}$ band filling.}
\label{F:rPhD}
\end{figure}

   Let us proceed to numerical investigations.
We draw ground-state phase diagrams by calculating $E_{\rm HF}$ in
the thermodynamic limit.
We show in Fig. \ref{F:PhD5} typical phase diagrams at $\frac{5}{6}$
band filling, which describe the as-grown MMX compounds.
As for CDW states, M-CDW is stabilized in the small-$\alpha$ region,
while it is mostly replaced by X-CDW with the increase of $\alpha$.
When the X$p_z$ orbital stays far bellow the M$d_{z^2}$ orbital, a
certain Coulomb repulsion on the X site is necessary for the
stabilization of X-CDW, which is accompanied by the alternating
charge densities on the X sublattice and therefore needs the driving
force $U_{\rm X}$ to push up the X$p_z$ electrons.
On the other hand, the SDW states are stabilized for large enough
$U_{\rm M}$ or $U_{\rm X}$.
Ni complexes, whose $U_{\rm M}$ is relatively large, may be candidates
for M-SDW.
Lower-lying X$p_z$ orbitals less stabilize X-SDW whose X$p_z$
orbitals are not fully occupied.
Fluctuations of the X-SDW type are interesting when
$\varepsilon_{\rm M}$ and $\varepsilon_{\rm X}$ are close enough.
The present observations are qualitatively summarized in Fig.
\ref{F:rPhD}.
$U_{\rm M}$ favors M-SDW, while $U_{\rm X}$ is advantageous to X-SDW.
The CDW states both dislike the on-site Coulomb repulsions,
especially $U_{\rm M}$.
M-CDW and X-CDW are almost equal with respect to $U_{\rm M}$,
$U_{\rm X}$, and $V_{\rm MX}$, but they are distinguished by
$V_{\rm MM}$.
Therefore, unless $\alpha$ is small enough, X-CDW is more stabilized.
With the increase of $\varepsilon_{\rm M}-\varepsilon_{\rm X}$, the
density waves on the X sublattice are generally reduced.
Within the mean-field approximation, the Stoner's ferromagnetism is
necessarily stabilized for large enough $U_{\rm M}$ and $U_{\rm X}$.

\begin{figure}
\begin{flushleft}
\mbox{\psfig{figure=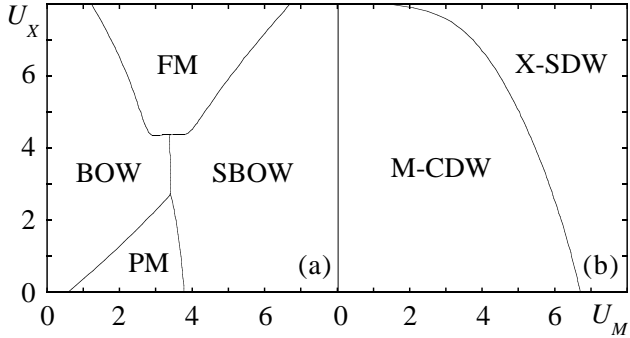,width=85mm,angle=0}}
\end{flushleft}
\vskip 0mm
\caption{Typical ground-state phase diagrams under hole doping, where
         $V_{\rm MM}=V_{\rm MX}=1.0$, $\alpha=0.9$, $\beta=0.6$, and
         $\varepsilon_{\rm M}-\varepsilon_{\rm X}=1.0$.
         (a) $\frac{4}{6}$ band filling.
         (b) $\frac{3}{6}$ band filling.}
\label{F:PhD43}
\end{figure}

\begin{figure}
$\!\!\!$\mbox{\psfig{figure=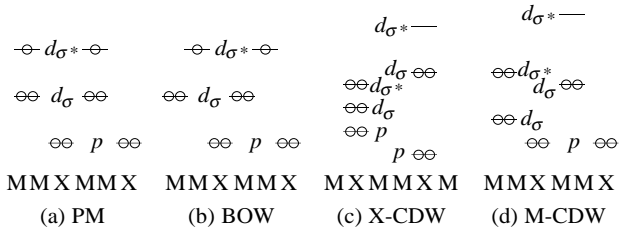,width=100mm,angle=0}}
\vskip 5mm
\caption{Schematic representation of the electronic structures of
         the MMX chain in the ground state of the nonmagnetic
         density-wave type in the case of neglecting any Coulomb
         correlation.}
\label{F:ES}
\end{figure}

   Finally, we mention the hole-doped system.
Typical phase diagrams at $\frac{4}{6}$ and $\frac{3}{6}$ band
fillings are shown in Fig. \ref{F:PhD43}.
Considering the nesting vector for the Fermi surfaces, it is
convincing that density waves of $q=\pi$ appear at $\frac{5}{6}$ and
$\frac{3}{6}$ band fillings, while they are less stabilized at
$\frac{4}{6}$ band filling.
In this context, let us consider the electronic structures, which are
illustrated in Fig. \ref{F:ES} at the idea of the same-valence metals
forming mixed orbitals \cite{Kita}.
The point is that the highest $d_{\sigma^*}$ orbitals are all empty
in any $q=\pi$ density-wave state, while they are occupied by finite
electrons in any $q=0$ density-wave state.
Under the doping of a hole per unit cell, $q=0$ states can closely
compete with $q=\pi$ states.
\vskip 0mm
\vskip 0mm

   Physical research on MMX compounds seems to be still in its early
stage.
The likely ground state of the X-CDW type in Pt$_2$(dta)$_4$I has
quite recently been supported by a single-band-model calculation
\cite{Kuwa} as well.
However, from the experimental point of view, further verification,
such as Raman spectroscopy and X-ray structural analysis, may be
provided.
The mobile bimetallic units are still wrapped in mystery.
We hope the present calculations will motivate and accelerate further
synthesis and measurements of these fascinating materials.

   It is a pleasure to thank Prof. H. Okamoto, Prof. K. Yonemitsu,
and Dr. M. Kuwabara for fruitful discussions.
This work was supported by the Japanese Ministry of Education,
Science, and Culture, and the Sanyo-Broadcasting Foundation for
Science and Culture.
The numerical calculation was done using the facility of the
Supercomputer Center, Institute for Solid State Physics, University
of Tokyo.

\widetext
\end{document}